\RequirePackage{fix-cm}
\documentclass[smallextended]{svjour3}       
\smartqed  
\usepackage{graphicx}
\usepackage{amsmath}
\usepackage{amssymb}
\usepackage{graphicx}
\usepackage{color}

\usepackage[mathscr,scaled=1.15]{urwchancal}
\DeclareFontFamily{OT1}{pzc}{}
\DeclareFontShape{OT1}{pzc}{m}{it}%
{<-> s * [1.15] pzcmi7t}{}
\DeclareMathAlphabet{\mathpzc}{OT1}{pzc}{m}{it}

\journalname{Few-Body Systems}

\begin{document}

\title{Nucleon viewed as a Borromean Bound-State
%
}


\author{Jorge Segovia \and C\'edric Mezrag \and \\ Lei Chang \and Craig D. Roberts}


\institute{Jorge Segovia (speaker) \at
           IFAE and BIST, Universitat Aut\`onoma de Barcelona, E-08193 Bellaterra (Barcelona), Spain \\           
           \email{jsegovia@ifae.es}
\and           
           C\'edric Mezrag \at
           Istituto Nazionale di Fisica Nucleare, Sezione di Roma, P. le A. Moro 2, I-00185 Roma, Italy
\and
           Lei Chang \at
           School of Physics, Nankai University, Tianjin 300071, China
\and
           Craig D. Roberts \at
           Physics Division, Argonne National Laboratory, Argonne, Illinois 60439, USA           
}

\date{Received: date / Accepted: date}

\maketitle

\begin{abstract}
We explain how the emergent phenomenon of dynamical chiral symmetry breaking ensures that Poincar\'e covariant analyses of the three valence-quark scattering problem in continuum quantum field theory yield a picture of the nucleon as a Borromean bound-state, in which binding arises primarily through the sum of two separate contributions. One involves aspects of the non-Abelian character of Quantum Chromodynamics that are expressed in the strong running coupling and generate tight, dynamical color-antitriplet quark-quark correlations in the scalar-isoscalar and pseudovector-isotriplet channels. This attraction is magnified by quark exchange associated with diquark breakup and reformation, which is required in order to ensure that each valence-quark participates in all diquark correlations to the complete extent allowed by its quantum numbers. Combining these effects, we arrive at a properly antisymmetrised Faddeev wave function for the nucleon and calculate, e.g. the flavor-separated versions of the Dirac and Pauli form factors and the proton's leading-twist parton distribution amplitude. We conclude that available data and planned experiments are capable of validating the proposed picture.
%
\end{abstract}


\newpage

\section{Introduction}
\label{sec:introduction}

The proton is the core of the hydrogen atom, lies at the heart of every nucleus, and has never been observed to decay; but it is nevertheless a composite object, whose properties and interactions are determined by its valence-quark content: $u$ + $u$ + $d$, \emph{i.e}.\ two up ($u$) quarks and one down ($d$) quark. So far as is now known~\cite{Patrignani:2016xqp}, bound-states seeded by two valence-quarks do not exist; and the only two-body composites are those associated with a valence-quark and -antiquark, \emph{i.e}.\ mesons. These features are supposed to derive from color confinement. Suspected to emerge in Quantum Chromodynamics (QCD), confinement is an empirical reality; but there is no universally agreed theoretical understanding~\cite{Roberts:2016vyn}.

Such observations lead one to a position from which the proton may be viewed as a Borromean bound-state, \emph{viz}.\ a system constituted from three bodies, no two of which can combine to produce an independent, asymptotic two-body bound-state. In QCD the complete picture of the proton is more complicated, owing, in large part, to the loss of particle number conservation in quantum field theory and the concomitant frame- and scale-dependence of any Fock space expansion of the proton's wave function~\cite{Dirac:1949cp,Keister:1991sb,Coester:1992cg,Brodsky:1997de}. Notwithstanding that, the Borromean analogy provides an instructive perspective from which to consider both quantum mechanical models and continuum treatments of the nucleon bound-state problem.  


\vspace*{-0.50cm}
\section{Diquark correlations}
\label{sec:diquarks}

Dynamical chiral symmetry breaking (DCSB) is another of QCD's emergent phenomena. It explains how a nearly massless current light quark acquires a momentum-dependent mass, namely the constituent quark mass, due to its interaction with the gluon medium; and delivers a mass-function whose value at infrared momenta explains $98\%$ of the proton mass. DCSB also ensures the existence of nearly-massless pseudo-Goldstone modes (pions) and, in the presence of these modes, no flux-tube between a static color source and sink can have a measurable existence. To verify this statement, consider such a tube being stretched between a source and sink. The potential energy accumulated within the tube may increase only until it reaches that required to produce a particle-antiparticle pair of the theory's pseudo-Goldstone modes. Simulations of lattice-regularised QCD (lQCD) show~\cite{Bali:2005fu,Prkacin:2005dc} that the flux-tube then disappears instantaneously along its entire length, leaving two isolated color-singlet systems. The length-scale associated with this effect in QCD is around $1/3\,\text{fm}$ and hence if any such string forms, it would dissolve well within hadron interiors.

Another equally important consequence of DCSB is less well known. Namely, any interaction capable of creating pseudo-Goldstone modes as bound-states of a light dressed-quark and -antiquark, and reproducing the measured value of their leptonic decay constants, will necessarily also generate strong color-antitriplet correlations between any two dressed quarks contained within a nucleon (and kindred baryons). Although a rigorous proof within QCD cannot be claimed, this assertion is based upon an accumulated body of evidence, gathered in two decades of studying two- and three-body bound-state problems in hadron physics (see, for instance, Refs.~\cite{Cahill:1987qr,Cahill:1988dx,Bender:1996bb,Maris:2002yu,Bhagwat:2004hn,Roberts:2011wy,Segovia:2013eca,Segovia:2014aza,Segovia:2015ufa,Segovia:2015hra,Eichmann:2016yit,Segovia:2016zyc,Burkert:2017djo}). No realistic counter examples are known; and the existence of such diquark correlations is also supported by lQCD~\cite{Alexandrou:2006cq,Babich:2007ah}.


\begin{figure}[!t]
\begin{center}
\hspace*{0.10cm}
\includegraphics[clip,width=0.55\textwidth,height=0.20\textheight]{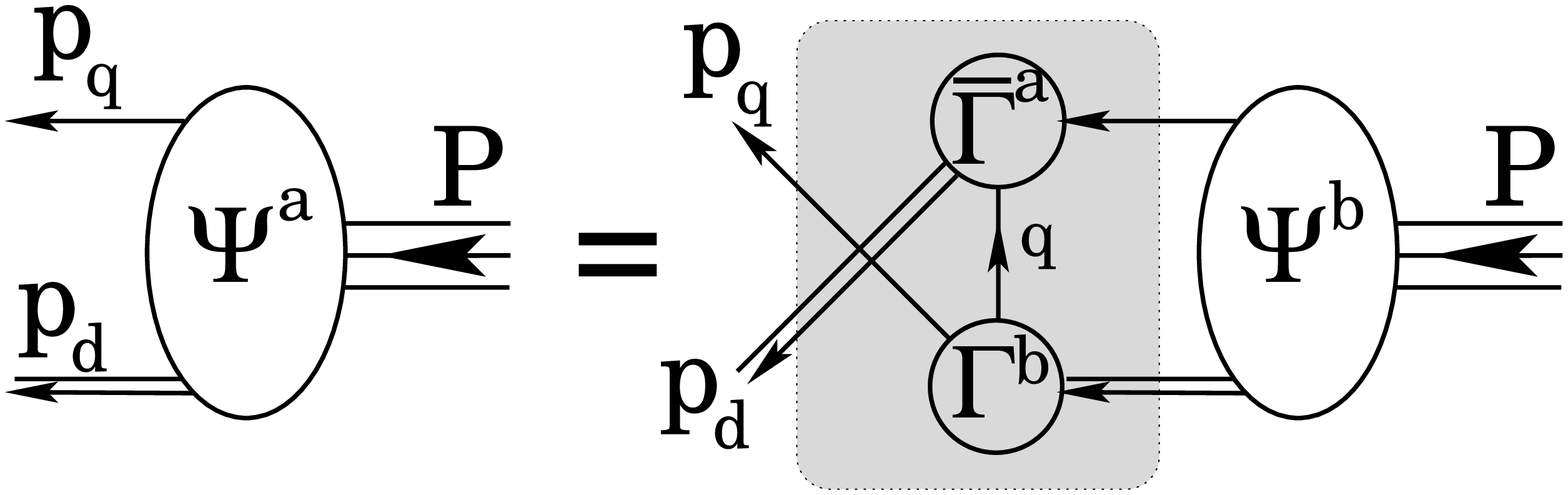}
\hspace*{0.20cm}
\includegraphics[clip,width=0.40\textwidth,height=0.22\textheight]{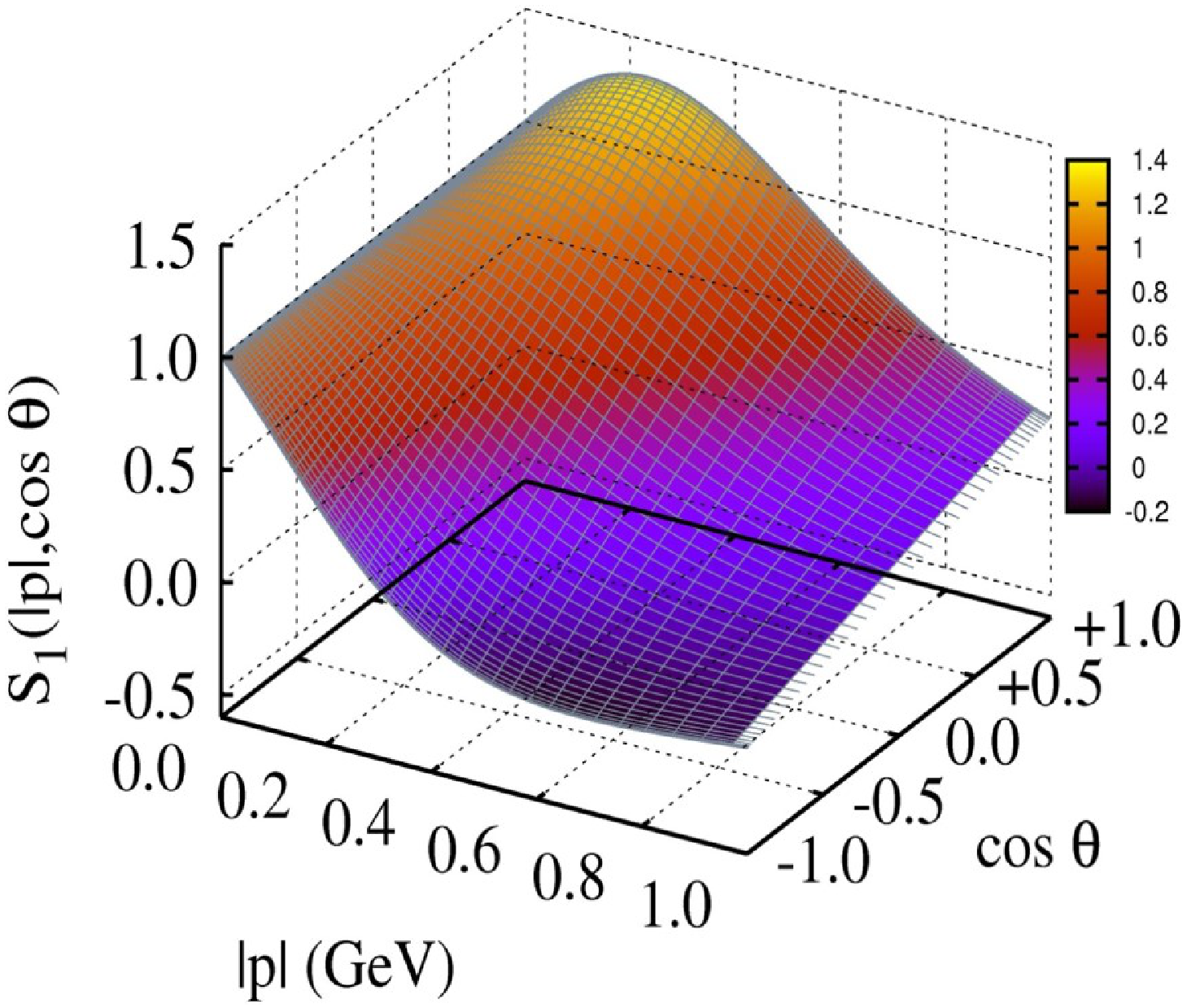}
\caption{\label{fig:Faddeev} {\it Left panel:} Poincar\'e covariant Faddeev equation. $\Psi$ is the Faddeev amplitude for a baryon of total momentum $P= p_q + p_d$, where $p_{q,d}$ are, respectively, the momenta of the quark and diquark within the bound-state. The shaded area demarcates the Faddeev equation kernel: {\it single line}, dressed-quark propagator; $\Gamma$, diquark correlation amplitude; and {\it double line}, diquark propagator.
{\it Right panel:} Dominant piece in the nucleon's eight-component Poincar\'e-covariant Faddeev amplitude: $S_1(|p|,\cos\theta)$. In the nucleon rest frame, this term describes that piece of the quark--scalar-diquark relative momentum correlation which possesses zero intrinsic quark--diquark orbital angular momentum, i.e. $L=0$ before the propagator lines are reattached to form the Faddeev wave function. We denote $p= P/3-p_q$ and $\cos\theta = p\cdot P/\sqrt{p^2 P^2}$. The amplitude is normalized such that its $U_0$ Chebyshev moment is unity at $|p|=0$.
\vspace*{-0.50cm}
}
\vspace*{-0.20cm}
\end{center}
\end{figure}

\vspace*{-0.50cm}
\section{Nucleon structure}
\label{sec:nucleon}

The existence of tight diquark correlations considerably simplifies analyses of the three valence-quark scattering problem and hence the baryon bound state, in particular the nucleon case, because it reduces that task to solving a Poincar\'e covariant Faddeev equation~\cite{Cahill:1988dx} depicted in the left panel of Fig.~\ref{fig:Faddeev}. The three-gluon vertex is not explicitly part of the bound-state kernel in this picture of the nucleon. Instead, one capitalizes on the fact that phase-space factors materially enhance two-body interactions over $n\geq 3$-body interactions and exploits the dominant role played by diquark correlations in the two-body subsystems. Then, whilst an explicit three-body term might affect fine details of nucleon structure, the dominant effect of non-Abelian multi-gluon vertices is expressed in the formation of diquark correlations. Such a nucleon is then a compound system whose properties and interactions are primarily determined by the quark$+$diquark structure evident in the left panel of Fig.~\ref{fig:Faddeev}.

A nucleon (and kindred baryons) described by the left panel of Fig.~\ref{fig:Faddeev} is a Borromean bound-state, the binding within which has two contributions. One part is expressed in the formation of tight diquark correlations. That is augmented, however, by attraction generated by the quark exchange depicted in the shaded area of the left panel of Fig.~\ref{fig:Faddeev}.  This  exchange ensures that diquark correlations within the nucleon are fully dynamical: no quark holds a special place because each one participates in all diquarks to the fullest extent allowed by its quantum numbers. The continual rearrangement of the quarks guarantees, \emph{inter} \emph{alia}, that the nucleon's dressed-quark wave function complies with Pauli statistics.

It is important to highlight that both scalar-isoscalar and pseudovector-isotriplet diquark correlations feature within a nucleon. The  relative probability of scalar versus pseudovector diquarks in a nucleon is a dynamical statement. Realistic computations predict a scalar diquark strength of approximately $62\%$~\cite{Segovia:2014aza,Segovia:2015ufa,Segovia:2015hra,Chen:2017pse}. As will become clear, this prediction can be tested by contemporary experiments.

The quark$+$diquark structure of the nucleon is elucidated in the right panel of Fig.~\ref{fig:Faddeev}, which depicts the leading component of its Faddeev amplitude: with the notation of Ref.~\cite{Segovia:2014aza}, $S_1(|p|,\cos\theta)$, computed using the Faddeev kernel described therein. This function describes a piece of the quark$+$scalar-diquark relative momentum correlation. Notably, in this solution of a realistic Faddeev equation there is strong variation with respect to both arguments. Support is concentrated in the forward direction, $\cos\theta >0$, so that alignment of $p$ and $P$ is favored; and the amplitude peaks at $(|p|\simeq M_N/6,\cos\theta=1)$, whereat $p_q \approx P/2 \approx p_d$ and hence the \emph{natural} relative momentum is zero. In the anti-parallel direction, $\cos\theta<0$, support is concentrated at $|p|=0$, \emph{i.e}.\ $p_q \approx P/3$, $p_d \approx 2P/3$.


\vspace*{-0.50cm}
\section{Nucleon current}
\label{sec:current}

The Poincar\'e-covariant photon-nucleon interaction current is:
\begin{equation}
J_{\mu}(K,Q) = i e \,\bar{u}(P_{f})\,\left[ \gamma_{\mu} F_{1}(Q^{2}) + \frac{\sigma_{\mu\nu}\,Q_{\nu}}{2m_{N}}\,F_{2}(Q^{2})\right] u(P_{i}) \,,
\label{eq:Jnucleon}
\end{equation}
where $P_{i}$ ($P_{f}$) is the momentum of the incoming (outgoing) nucleon; $Q=P_{f} - P_{i}$, $K=(P_{i}+P_{f})/2$: for elastic scattering, $K\cdot Q=0$, $K^{2} = - m_{N}^{2} (1+\tau_{N})$, $\tau_{N} = Q^{2}/(4 m_{N}^{2})$. The functions $F_{1,2}$ are, respectively, the Dirac and Pauli form factors: $F_1(0)$ expresses the bound-state's electric charge and $F_2(0)$ its anomalous magnetic moment, $\kappa_{N=n,p}$. The Sachs electric and magnetic form factors are, respectively, $G_E = F_1 - \tau_N F_2$ and $G_M = F_1+F_2$.

\begin{figure}[!t]
\begin{center}
\includegraphics[clip,width=0.45\textwidth,height=0.25\textheight]{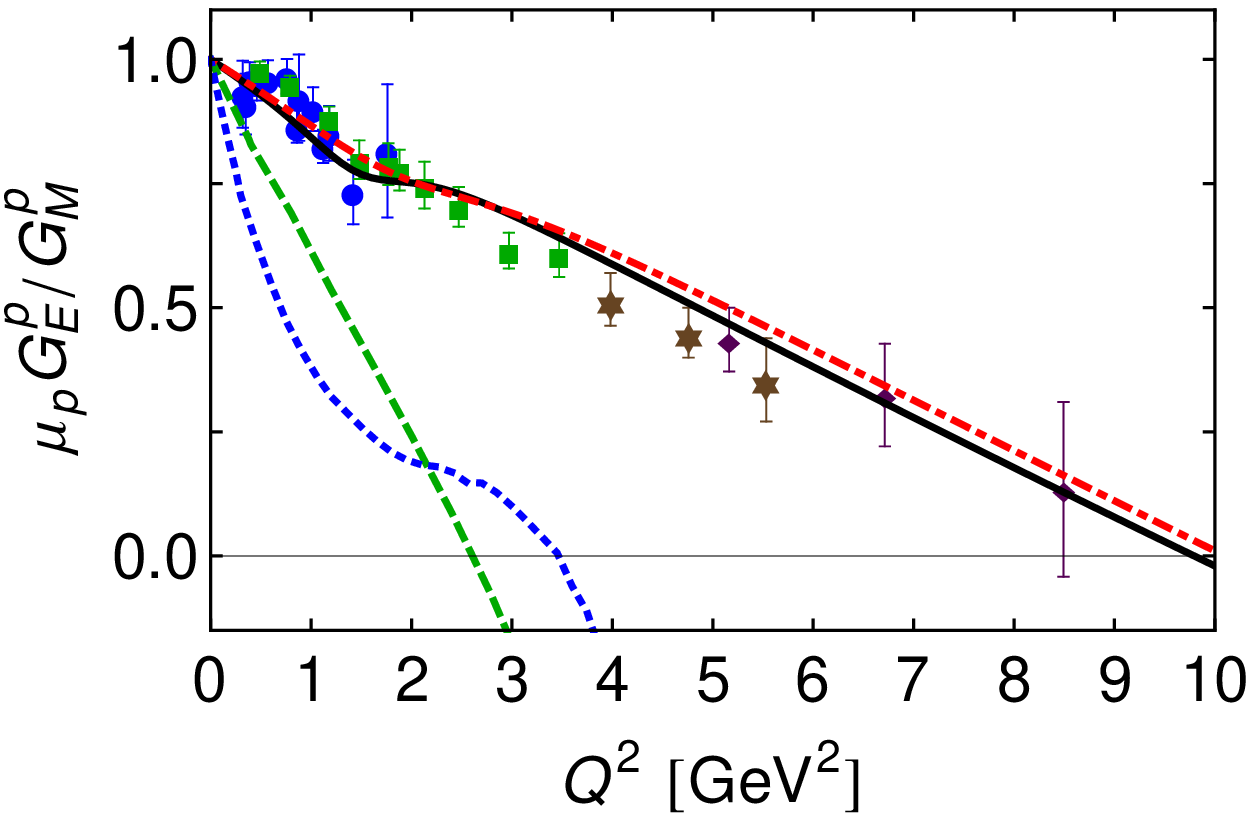}
\hspace*{0.50cm}
\includegraphics[clip,width=0.45\textwidth,height=0.25\textheight]{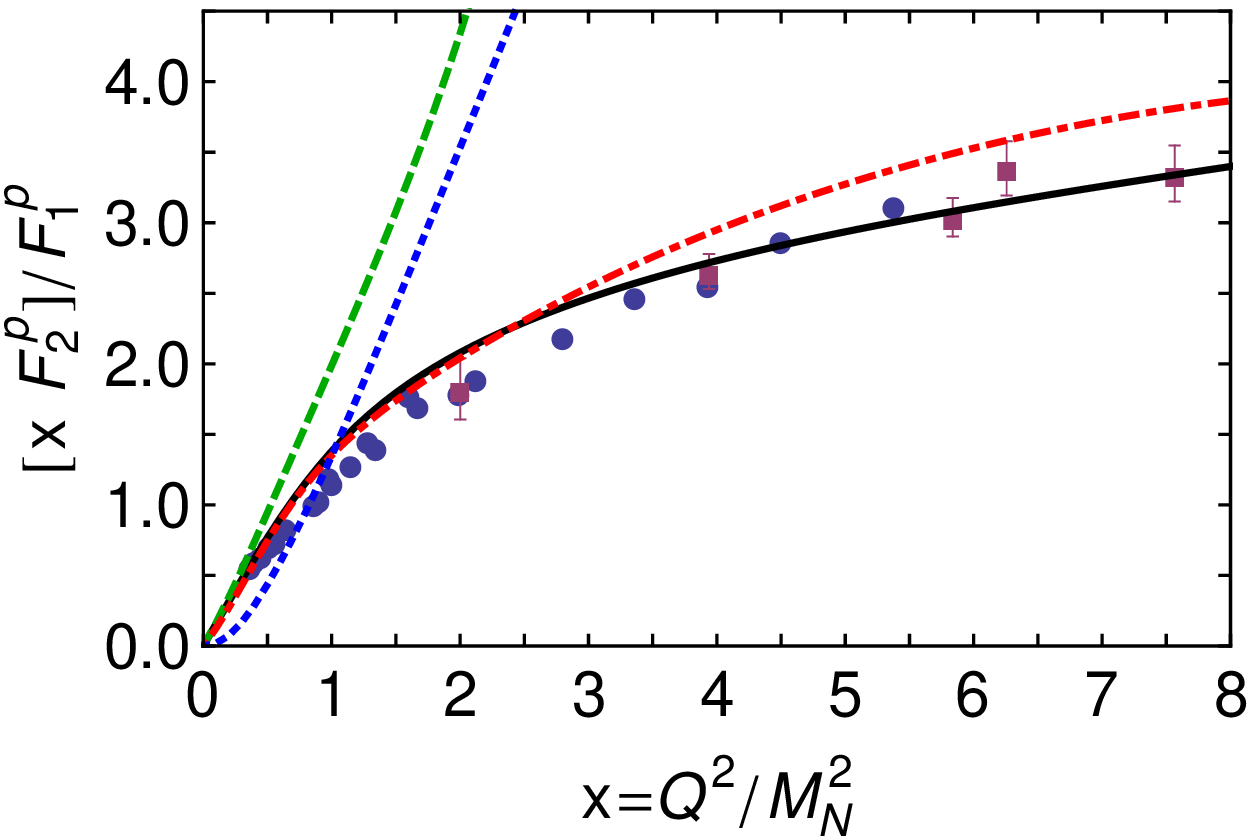}
\caption{\label{figGEGM} 
{\it Left panel:} Computed ratio of proton electric and magnetic form factors. Curves: solid (black) -- full result, determined from the complete proton Faddeev wave function and current; dot-dashed (red) -- momentum-dependence of scalar-diquark contribution; dashed (green) -- momentum-dependence produced by that piece of the scalar diquark contribution to the proton's Faddeev wave function which is purely $S$-wave in the rest-frame; dotted (blue) -- momentum-dependence of pseudovector diquark contribution.
All partial contributions have been renormalized to produce unity at $Q^2=0$.
Data: circles (blue) \cite{Gayou:2001qt}; squares (green)  \cite{Punjabi:2005wq}; asterisks (brown) \cite{Puckett:2010ac}; and diamonds (purple) \cite{Puckett:2011xg}.
{\it Right panel:} Proton ratio $R_{21}(x) = x F_2(x)/F_1(x)$, $x=Q^2/M_N^2$. The legend is as above. Experimental data taken from Ref.~\protect\cite{Cates:2011pz}.
\vspace*{-0.50cm}
}
\end{center}
\end{figure}

Left panel of Fig.~\ref{figGEGM} depicts the ratio of proton electric and magnetic form factors, $R_{EM}(Q^2)=\mu_p G_E(Q^2)/G_M(Q^2)$ $\mu_p=G_M(0)$. A series of experiments~\cite{Gayou:2001qt,Punjabi:2005wq,Puckett:2010ac,Puckett:2011xg} has determined that $R_{EM}(Q^2)$ decreases almost linearly with $Q^2$ and might become negative for $Q^2 \gtrsim 8\,\text{GeV}^2$. A clear conclusion from the left panel of Fig.~\ref{figGEGM} is that pseudovector diquark correlations have little influence on the momentum dependence of $R_{EM}(Q^2)$. Their contribution is indicated by the dotted (blue) curve, which was obtained by setting the scalar diquark component of the proton's Faddeev amplitude to zero and renormalizing the result to unity at $Q^2=0$. As apparent from the dot-dashed (red) curve, the evolution of $R_{EM}(Q^2)$ with $Q^2$ is primarily determined by the proton's scalar diquark component. However, the dashed (green) curve reveals something more, \emph{i.e}.\ components of the nucleon associated with quark-diquark orbital angular momentum $L=1$ in the nucleon rest frame are critical in explaining the data. Notably, the presence of such components is an inescapable consequence of solving a realistic Poincar\'e-covariant Faddeev equation.

The right panel of Fig.~\ref{figGEGM} shows the proton's ratio: $R_{21}(x) = x F_2(x)/F_1(x)$, with $x=Q^2/M_N^2$. As in the $R_{EM}$ case, the momentum dependence of $R_{21}(x)$ is principally determined by the scalar diquark component of the proton and the rest-frame $L=1$ terms are again seen to be critical in explaining the data.

The Hall A Collaboration of JLab~\cite{Riordan:2010id} released in $2011$ precise data of the neutron's electric form factor up to $Q^{2}=3.4\,{\rm GeV}^{2}$. This allowed, for the first time, an analysis of the $u$- and $d$-quark contributions of the Dirac and Pauli form factors of the proton on a sizable domain of momentum transfer~\cite{Cates:2011pz}.

\begin{figure}[!t]
\begin{center}
\includegraphics[clip,width=0.45\textwidth,height=0.25\textheight]{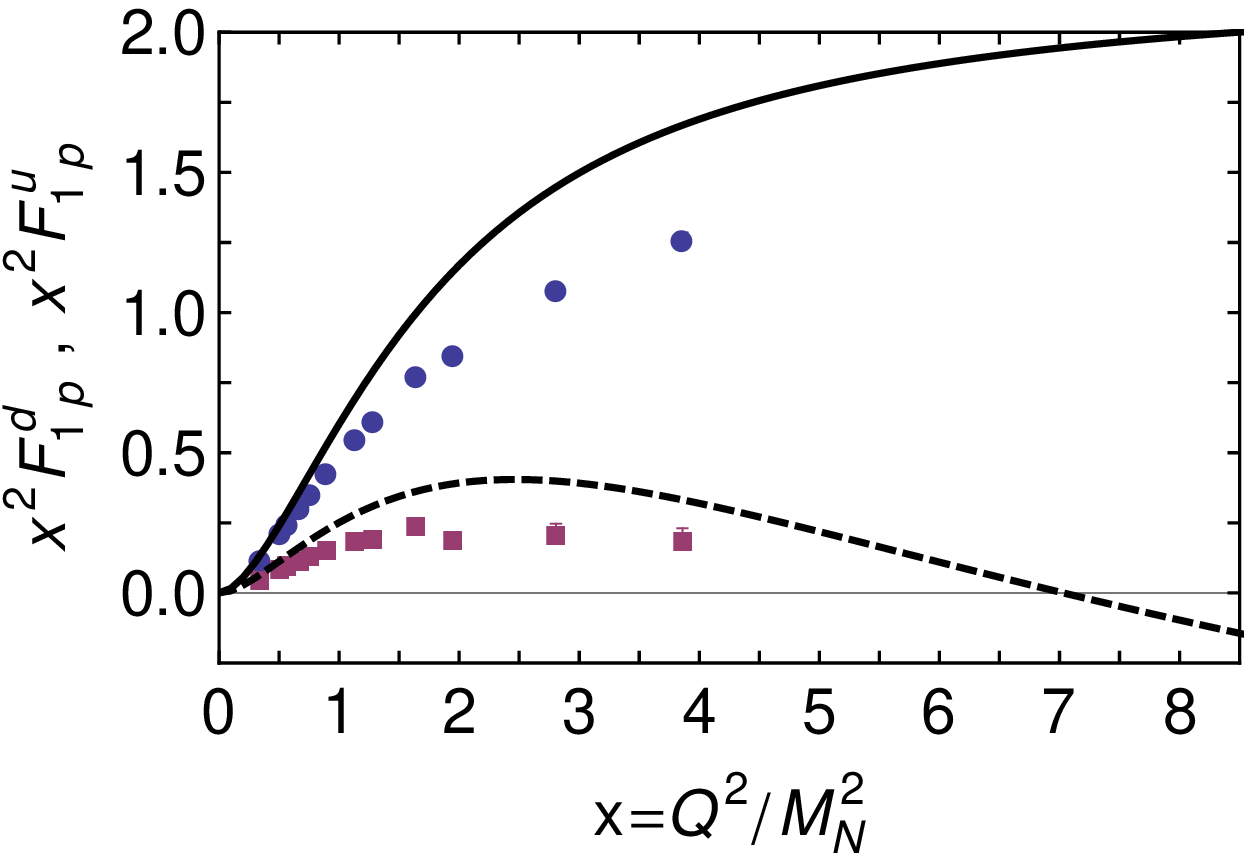}
\hspace*{0.50cm}
\includegraphics[clip,width=0.45\textwidth,height=0.25\textheight]{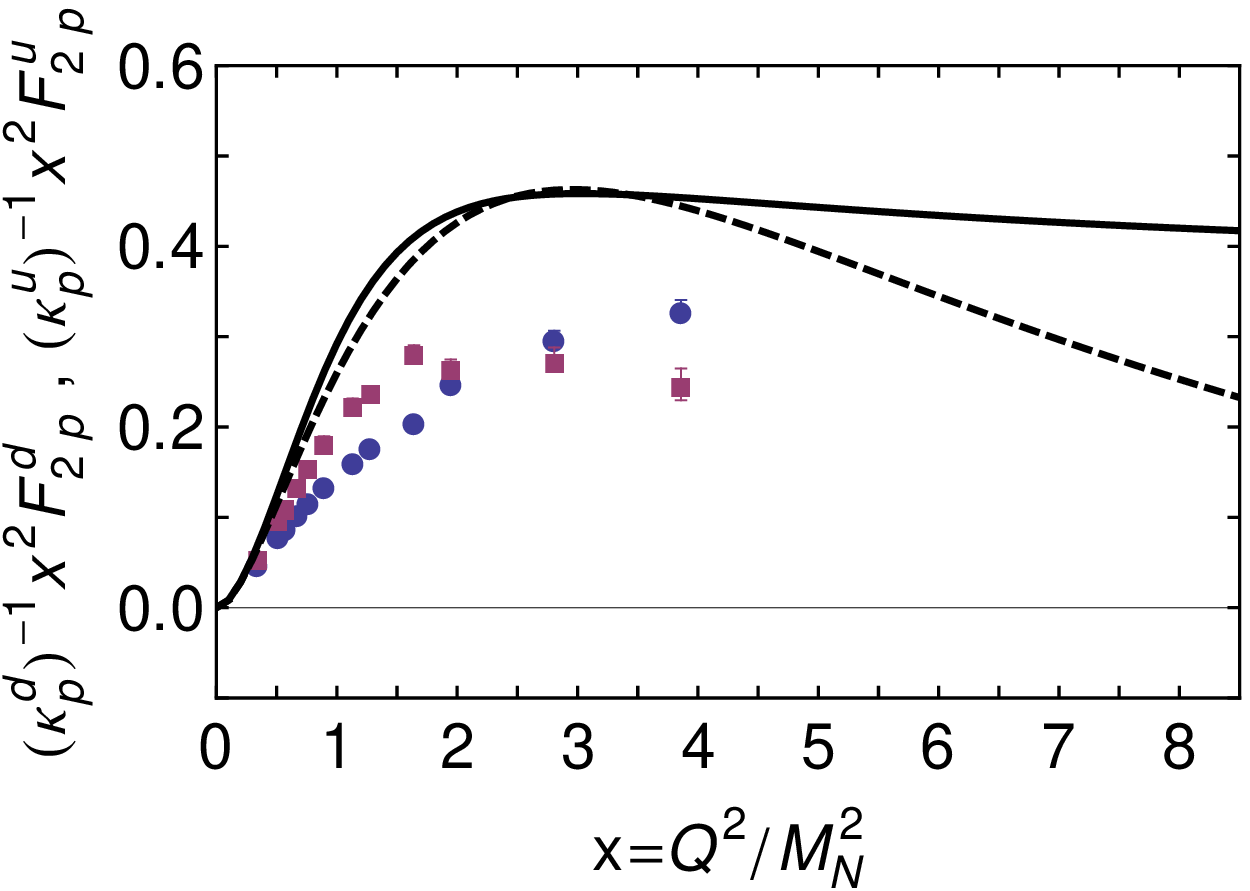}
\caption{\label{fig:F1F2fla1} \emph{Left panel}. Flavor separation of the proton's Dirac form factor as a function of $x=Q^2/M_N^2$. Curves: solid -- $u$-quark; and dashed $d$-quark contribution.  Data: circles -- $u$-quark; and squares -- $d$-quark.
\emph{Right panel}.  Same for Pauli form factor.
Data: Refs.\,\protect\cite{Zhu:2001md,Bermuth:2003qh,Warren:2003ma,Glazier:2004ny,Plaster:2005cx,Riordan:2010id,Cates:2011pz}.
\vspace*{-0.50cm}
}
\end{center}
\end{figure}

Figure~\ref{fig:F1F2fla1} displays the proton's flavor separated Dirac and Pauli form factors. The salient features of the data are: (i) the $d$-quark contribution to $F_1^p$ is far smaller than the $u$-quark contribution; (ii) $F_2^d/\kappa_d>F_2^u/\kappa_u$ on $x<2$ but this ordering is reversed on $x>2$; and (iii) in both cases the $d$-quark contribution falls dramatically on $x>3$ whereas the $u$-quark contribution remains roughly constant. Our calculations are in semi-quantitative agreement with the empirical data. They reproduce the qualitative behaviour and also predict a zero in $F_1^d$ at $x\simeq 7$. The zero in $F_1^d$ is a measure of the relative probability of finding pseudovector and scalar diquarks in the proton: with all other things held equal, the zero moves toward $x=0$ as the probability of finding a pseudovector diquark within the proton increases.

It is natural to seek an explanation for the pattern of behaviour in Fig.~\ref{fig:F1F2fla1}. We have emphasized that the proton contains scalar and pseudovector diquark correlations. The dominant piece of its Faddeev wave function is $u[ud]$; namely, a $u$-quark in tandem with a $[ud]$ scalar correlation, which produces $62\%$ of the proton's normalization~\cite{Segovia:2014aza,Chen:2017pse}. If this were the sole component, then photon--$d$-quark interactions within the proton would receive a $1/x$ suppression on $x>1$, because the $d$-quark is sequestered in a soft correlation, whereas a spectator $u$-quark is always available to participate in a hard interaction.  At large $x=Q^2/M_N^2$, therefore, scalar diquark dominance leads one to expect $F^d \sim F^u/x$. Available data are consistent with this prediction but measurements at $x>4$ are necessary for confirmation.

\vspace*{-0.50cm}
\section{Nucleon parton distribution amplitude}

Extreme challenges are faced when the constituents of a hadron are light. Most of these difficulties are related with the fact that particle-number is not conserved and thus a probability interpretation of the hadron's wave function is lost. This can be circumvented by using a light-front formulation because eigenfunctions of the Hamiltonian are then independent of the system's four-momentum~\cite{Keister:1991sb,Brodsky:1997de}.

The light-front wave function of a hadron with momentum $P$ and spin $\lambda$, $\Psi(P,\lambda)$, is complicated. In terms of perturbation theory's partons, $\Psi(P,\lambda)$ has a countably-infinite Fock-space expansion, with the $N$-parton term depending on $3N$ momentum variables, constrained such that their sum yields $P$, with a similar constraint on their spin (and flavor). Fortunately, collinear factorization in the treatment of hard exclusive processes entails that much can be gained merely by studying hadron leading-twist parton distribution amplitudes (PDAs)~\cite{Lepage:1979zb,Efremov:1979qk,Lepage:1980fj}. Such a PDA is obtained from the simplest term in the Fock-space expansion, \emph{e.g.} meson, quark-antiquark $(N=2)$ or baryon, three-quark $(N=3)$, with the constituents' light-front transverse momenta integrated out to a given scale, $\zeta$.

In the isospin-symmetry limit, the proton possesses one independent leading-twist (twist-three) PDA~\cite{Braun:2000kw}, denoted $\phi([x];\zeta)$ herein:
\begin{equation}
\begin{split}
\langle 0| \epsilon^{abc} \tilde{u}_{+}^{a}(z_1) C^{\dagger} & \gamma\cdot n \tilde{u}_{-}^{b}(z_2) \gamma\cdot n \tilde{d}_{+}^{c}(z_3) |P,+ \rangle =: \frac{1}{2} f_p \\
&
\times (n\cdot P) (\gamma\cdot n) N_+ \int d[x] \phi([x];\zeta) e^{-in\cdot P \sum_i x_i z_i}
\end{split}
\end{equation}
where $n^2=0$; $(a,b,c)$ are color indices; $\psi_{\pm}=H_{\pm}\psi := (1/2)(I_D \pm \gamma_5)\psi$; $\tilde{q}$ indicates matrix transpose; $C$ is the charge conjugation matrix, $N=N(P)$ is the proton's Euclidean Dirac spinor; $\int [dx] = \int_0^1 dx_1 dx_2 dx_3 \delta(1-\sum_i x_i)$; and $f_p$ measures the proton's ``wave function at the origin''.

\begin{figure}[!t]
\begin{center}
\includegraphics[clip, trim={0.0cm 0.4cm 0.0cm 2.6cm}, width=0.45\textwidth, height=0.225\textheight]{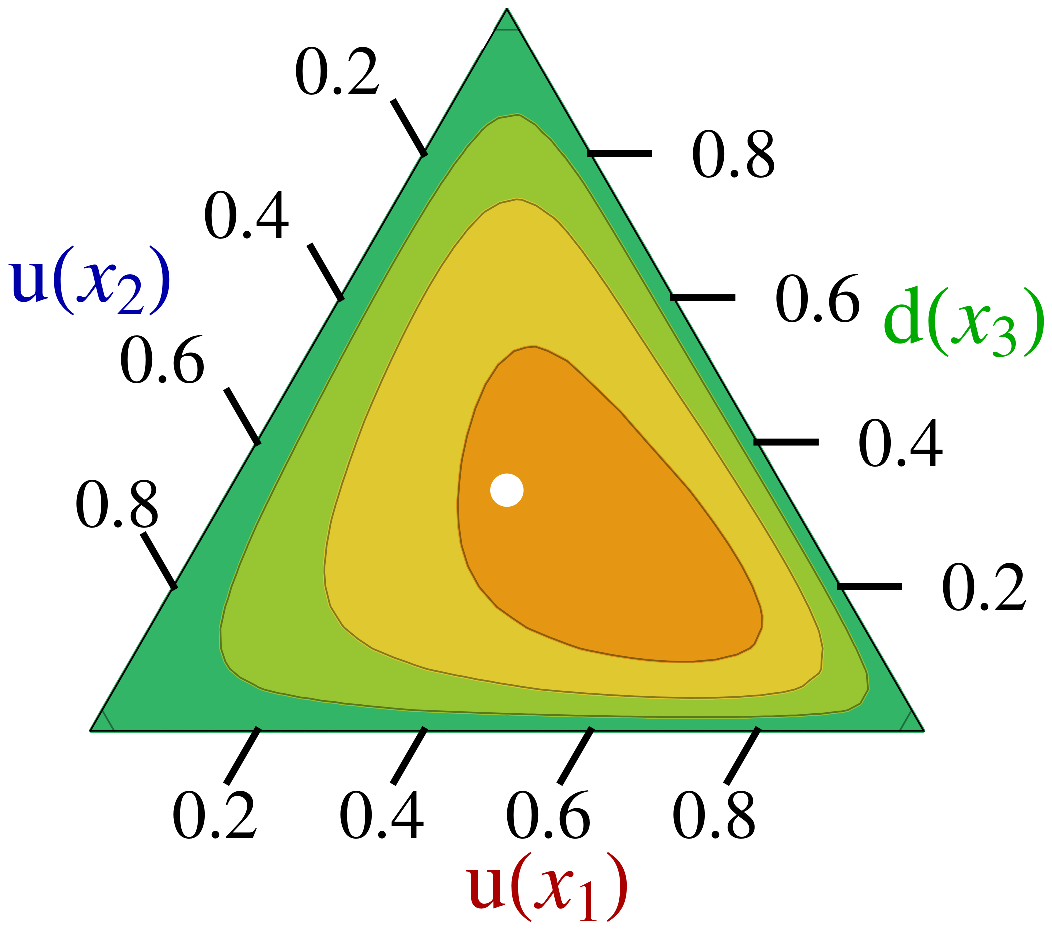}
\hspace*{0.10cm}
\includegraphics[clip, trim={0.0cm 0.4cm 0.0cm 2.0cm}, width=0.520\textwidth, height=0.225\textheight]{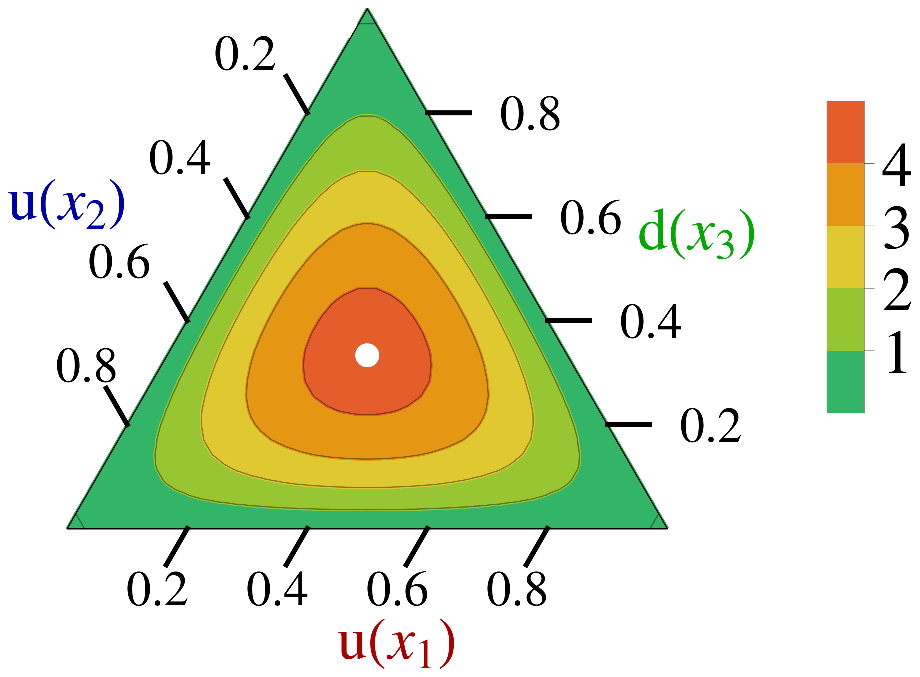}
%
\caption{\label{fig:PDA} Barycentric plots: \emph{left panel} -- computed proton PDA evolved to $\zeta=2\,\text{GeV}$, which peaks at $([x])=(0.55, 0.23, 0.22)$; \emph{right panel} -- conformal limit PDA, $\phi_N^{cl}([x]) = 120 x_1 x_2 x_3$.
\vspace*{-0.50cm}
}
\end{center}
\end{figure}

The proton's leading-twist PDA, $\phi([x])$, can be computed once the Poincar\'e-covariant wave function of the proton is in hand. In Ref.~\cite{Mezrag:2017znp}, we have used simple perturbation theory integral representations (PTIRs) for all elements in the Faddeev wave function depicted in Fig.~\ref{fig:Faddeev}, therewith defining models constrained by the best available solutions of the continuum three-valence-body bound-state equations. Figure~\ref{fig:PDA} shows our calculation of the leading-twist PDA of the ground-state nucleon and compares it with QCD's asymptotic limit. One can see that $\phi([x])$ is broader than $\phi_N^{cl}([x])$ and decreases monotonically away from its maximum in all directions. Moreover, the observed asymmetry is due to the presence and relative weight of scalar and pseudovector diquark correlations inside the proton: the presence of pseudovector correlations is essential.

\vspace*{-0.50cm}
\section{Summary}
\label{sec:summary}

We explained how the emergent phenomenon of dynamical chiral symmetry breaking ensures that Poincar\'e covariant analyses of the three valence-quark scattering problem in continuum quantum field theory yield a picture of the nucleon as a Borromean bound-state, in which binding arises primarily through the sum of two separate contributions. One involves aspects of the non-Abelian character of QCD that are expressed in the strong running coupling and generate tight, dynamical color-antitriplet quark-quark correlations in the isoscalar scalar and isovector-pseudovector channels. This attraction is magnified by quark exchange associated with diquark breakup and reformation, which is required in order to ensure that each valence-quark participates in all diquark correlations to the complete extent allowed by its quantum numbers.


\vspace*{-0.20cm}
\begin{acknowledgements}
Work supported by: European Union's Horizon 2020 research and innovation programme under the Marie Sk\l{}odowska-Curie grant agreement no. 665919; Spanish MINECO's Juan de la Cierva-Incorporaci\'on programme with grant agreement no. IJCI-2016-30028; Spanish Ministerio de Econom\'ia, Industria y Competitividad under contract nos. FPA2014-55613-P and SEV-2016-0588; the Chinese Government's Thousand Talents Plan for Young Professionals; and U.S. Department of Energy, Office of Science, Office of Nuclear Physics, under contract no. DE-AC02-06CH11357.
\end{acknowledgements}

\vspace*{-0.50cm}

\bibliographystyle{spphys}       
\bibliography{Crit17}   

%
%
%

\end{document}